\documentclass[aps,prb,
 amsmath,amssymb,
 twocolumn, groupedaddress,
 showpacs]{revtex4-1}
\usepackage{bm}
\usepackage{graphicx}
\usepackage{gensymb}
\usepackage{amsmath}
\usepackage{amssymb}
\usepackage{color}
\usepackage{ulem}
\usepackage{here}

\begin{document}

\title{Unusual change in the Dirac-cone energy band upon two-step magnetic transition in CeBi}
\author{Hikaru Oinuma,$^1$ Seigo Souma,$^{2,3}$ Kosuke Nakayama,$^1$ Koji Horiba,$^4$ Hiroshi Kumigashira,$^{4,5}$ Makoto Yoshida,$^6$ Akira Ochiai,$^1$ Takashi Takahashi,$^{1,2,3}$ and Takafumi Sato$^{1,2,3}$}

\affiliation{$^1$Department of Physics, Tohoku University, Sendai 980-8578, Japan\\
$^2$Center for Spintronics Research Network, Tohoku University, Sendai 980-8577, Japan\\
$^3$WPI Research Center, Advanced Institute for Materials Research, Tohoku University, Sendai 980-8577, Japan\\
$^4$Institute of Materials Structure Science, High Energy Accelerator Research Organization (KEK), Tsukuba, Ibaraki 305-0801, Japan\\
$^5$Institute of Multidisciplinary Research for Advanced Materials (IMRAM), Tohoku University, Sendai 980-8577, Japan\\
$^6$Max-Planck-Institute for Solid State Research, Heisenbergstrsse 1, 70569 Stuttgart, Germany
}

\date{\today}

\begin{abstract}
We have performed angle-resolved photoemission spectroscopy (ARPES) on CeBi which undergoes a two-step antiferromagnetic (AF) transition with temperature. Soft-x-ray ARPES has revealed the inverted band structure at the {\it X} point of bulk Brillouin zone for CeBi (and also for LaBi) as opposed to LaSb with non-inverted band structure. Low-energy ARPES on CeBi has revealed the Dirac-cone band at the $\bar{\Gamma}$ point in the paramagnetic phase associated with the bulk band inversion. On the other hand, a double Dirac-cone band appears on entering the first AF phase at {\it T} = 25 K, whereas a single Dirac-cone band recovers below the second AF transition at {\it T} = 14 K. The present result suggests an intricate interplay between antiferromagnetism and topological surface states in CeBi.
\end{abstract}

\pacs{71.20.-b, 73.20.At, 79.60.-i}

\maketitle

\section{INTRODUCTION}

One of intriguing challenges in heavy-fermion physics is to understand the relationship between magnetism and electronic structure associated with the interplay of conduction and {\it f} electrons. Cerium monopnictides CeX$_p$ (X$_p$ = P, As, Sb, and Bi) are a typical example of heavy-fermion systems showing an exotic and complicated magnetic phase diagram as a function of temperature, pressure, and magnetic field, known as a ``devil's staircase" [\onlinecite{RosMig1, RosMig2, RosMig3, Kohgi2000}]. CeX$_p$ is a Kondo semimetal with the rock-salt structure having two hole pockets at the $\Gamma$ point in the bulk fcc Brillouin zone (BZ) arising from the X$_p$ {\it p} orbital, together with an electron pocket at the {\it X} point originating from the Ce 5{\it d} orbital [see Fig. 1(a)] [\onlinecite{Hasegawa}]. Ce 4{\it f} electrons are nearly localized in the crystal lattice (Ce$^{3+}$) and thereby the overall electronic structure of CeX$_p$ is essentially the same as that of LaX$_p$ with no 4{\it f} electrons [\onlinecite{Hasegawa}, \onlinecite{Settai1994}].

Unlike the case of RKKY(Ruderman-Kittel-Kasuya-Yosida)-type magnetic order in rare-earth metals, the complex magnetic properties of CeX$_p$, in particular CeSb, is explained in terms of the {\it p-f} mixing [\onlinecite{Kasuya_pf}] which is based on the strong hybridization between the X$_p$ {\it p} state and the crystal-field-split level of the Ce 4{\it f} state [\onlinecite{CF}], as supported by the observation of an energy shift of the {\it p} band in angle-resolved photoemission spectroscopy (ARPES) of CeSb [\onlinecite{Kumi1997, Ito2004, Takayama2009, SJang2019}] and the change in the Fermi surface in the de Haas-van Alphen experiment [\onlinecite{Kitazawa1988}] across the magnetic transition. As a result of strong hybridization, the Ce 4{\it f} state shows ferromagnetic coupling within the layer and leads to the complicated magnetic structure stabilized by a subtle balance between the entropy and the interlayer exchange coupling [\onlinecite{Boehm1979, Nakanishi1989, Kasuya1990}]. Such {\it p-f} mixing and the electron correlation are thought to be responsible for the observed anomalous behavior in the electrical transport and lattice parameters [\onlinecite{Okayama1992, Kasuya1993, Kohgi2000, Iwasa2002}] as well as the rich magnetic properties, making the physics of CeX$_p$ particularly fertile.  

Recently, Zeng {\it et al}. predicted from the first-principles band-structure calculations [\onlinecite{ZengArXiv}] that LaX$_p$ becomes a topological insulator due to the band inversion at the {\it X} point of bulk BZ. This report has renewed the interest for RX$_p$ (R = La and Ce) systems in topological aspects and triggered intensive theoretical and experimental investigations [\onlinecite{SJang2019, PJGuo2016, PJGuo2017, Stepanov2015, TaftiNP2016, SunNJP2016, KumarPRB2016, TaftiPNAS2016, Kumar2017, TaftiPRB2017, SinghaarXiv2017, FWu2017, CGuo2017, Joe2018, YWang2018B, YWang2018A, ZLi2017, ZengPRL2016, WuPRB2016, HasanArxiv2016, NiuPRB2016, JHe2016, NayakNC2017, Oinuma2017, LouPRB2017, YWu2017, Kuroda2018, BFeng2018, PLi2018}], resulting in the discovery of extremely large magnetoresistance and unusual resistivity plateau in LaSb and LaBi [\onlinecite{Stepanov2015, TaftiNP2016, SunNJP2016, KumarPRB2016, TaftiPNAS2016, Kumar2017, TaftiPRB2017, SinghaarXiv2017}], as well as the observation of Dirac-cone-like energy band in some RX$_p$ compounds [\onlinecite{WuPRB2016, HasanArxiv2016, NiuPRB2016, JHe2016, NayakNC2017, Oinuma2017, LouPRB2017, YWu2017, Kuroda2018, BFeng2018, PLi2018}]. While RSb was turned out to be topologically trivial [\onlinecite{ZengPRL2016, JHe2016, Oinuma2017, Kuroda2018}], ARPES studies of LaBi [\onlinecite{ZengPRL2016, HasanArxiv2016, NiuPRB2016, NayakNC2017, LouPRB2017}] have commonly revealed the Dirac-cone-like energy band of topological origin at the $\bar{\Gamma}$ and $\bar{M}$  point of the surface BZ, associated with the band inversion at the {\it X} point [note that the $\bar{M}$ point is equivalent to the projected bulk {\it X} point; see Fig. 1(a)]. 

Taking into account such magnetic and topological aspects of RX$_p$ family, one would naturally expect that CeBi is a unique candidate to study the interplay between magnetism and topological properties, since it shows interesting magnetic phases characterized by a two-step antiferromagnetic (AF) transition at {\it T} = 25 and 14 K under zero-magnetic field [\onlinecite{RosMig3}, \onlinecite{Kohgi2000}], in addition to the expected topological nature. In a broader perspective, it is of great importance to experimentally clarify the role of antiferromagnetism to the topological properties, which is currently a target of intensive debates in theories [\onlinecite{MongPRB2010, HGuo2011, CFang2013, Miyakoshi2013, Yoshida2013, Muller2014, CXLiu2014, RXZhang2015, CFang2015, JYu2017, Brzezicki2017, NHao2017, KWChang2018}] while no concrete experimental data have been hitherto reported.

In this article, we report the soft-x-ray (SX) and vacuum-ultraviolet (VUV) ARPES study of CeBi. We found that the paramagnetic phase of CeBi is characterized by a single Dirac-cone energy band at the $\bar{\Gamma}$  point originating from the band inversion at the {\it X} point in bulk BZ, whereas the energy dispersion of the Dirac-cone band is strongly reconstructed in the AF phase; surprisingly it also depends on the types of magnetic structures. We discuss the implications of our observation in relation to the magnetic band folding and the symmetry of the AF phase.

\section{EXPERIMENTAL}

CeBi single crystal and its reference materials LaBi and LaSb were grown by the Bridgman method with a tungsten heater furnace. High-purity starting materials of La/Ce (4N) and Sb/Bi (6N) with the ratio of 1:1.005 were sealed in a tungsten crucible using an electron beam welder. The crucible was heated above their melting points and then slowly pulled down from the heater. The obtained crystals were characterized by the x-ray diffraction measurements. SX-ARPES measurements were performed with a Scienta-Omicron SES2002 electron analyzer with energy-tunable synchrotron light at BL2 in Photon Factory (PF), KEK. We used linearly polarized light (horizontal polarization) of 400-630 eV. VUV-ARPES measurements were performed with a MBS A1 electron analyzer with the Xe discharge lamp at Tohoku University. We used the Xe-I line (8.437 eV). The energy resolutions for SX- and VUV-ARPES measurements were set to be 150 and 5 meV, respectively. Samples were cleaved {\it in situ} in an ultrahigh vacuum of $\sim$ 1$\times$10$^{-10}$ Torr along the (100) crystal plane. The Fermi level ($E_{\rm F}$) of samples was referenced to that of a gold film evaporated onto the sample holder.

\begin{figure}
\begin{center}
\includegraphics[width=3 in]{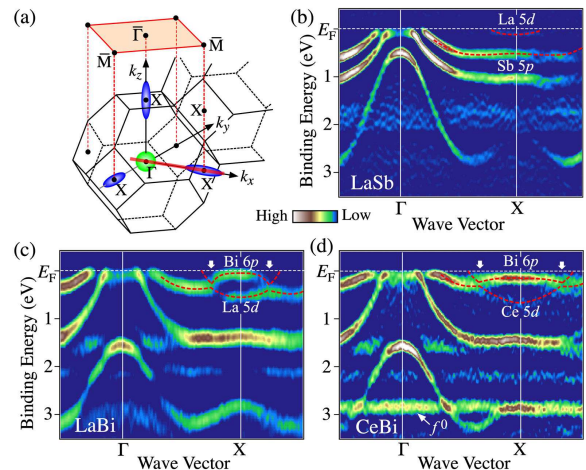}
  \hspace{0.2in}
\caption{(color online). (a) Bulk and surface BZs of RX$_p$ together with the $k$ cut (red solid line) where the ARPES data for (a)-(c) were obtained. (b) Second-derivative-intensity plot of EDC for LaSb at {\it T} = 30 K measured along the $\Gamma${\it X} cut in bulk BZ with SX photons. (c), (d) Same as (b) but for LaBi and CeBi, respectively. We estimated the photon energy that traces the  $\Gamma${\it X} cut to be $h\nu$ = 510, 500, and 505 eV, for LaSb, LaBi, and CeBi, respectively, by the normal-emission ARPES measurements. Red dashed curves in (a)-(c) are guides for the eyes to trace the band dispersion. White arrows in (c) and (d) highlight the intersection of Bi 6{\it p} and La/Ce 5{\it d} bands. Non-dispersive feature at $E_{\rm B}$ $\sim$ 2.9 eV in (d) is ascribed to the Ce 4$f^0$ final state [\onlinecite{Franciosi1981}, \onlinecite{Allen1981}].
}
\end{center}
\end{figure}

\section{RESULTS AND DISCUSSION}

First, we present the overall energy band structure of the bulk valence band. To discuss the electronic states of CeBi in terms of the topology, it is essential to clarify (i) the role of spin-orbit coupling (SOC) and (ii) the influence of Ce 4{\it f} electrons to the energy bands. Since the bulk-band inversion at the {\it X} point is thought to be directly linked to the topological nature of this system [\onlinecite{ZengArXiv, ZengPRL2016, HasanArxiv2016, NiuPRB2016, JHe2016, NayakNC2017, LouPRB2017, Oinuma2017, Kuroda2018, PLi2018}], we experimentally determined the bulk-band structure passing through the {\it X} point (along the $\Gamma${\it X} cut) of bulk 3D BZ [see Fig. 1(a)] with SX photons ($h\nu$ $\sim$ 500 eV), and compa the results between LaSb and LaBi to address the issue (i) since the replacement of Sb with Bi leads to a stronger SOC due to the heavier atomic mass of Bi. In addition, we compared LaBi ($f^0$ system) and CeBi ($f^1$ system) to address the issue (ii). The result for LaSb shown in Fig. 1(b) signifies an electronlike La 5{\it d} band crossing $E_{\rm F}$ around the {\it X} point and several holelike Sb 5{\it p} bands at the $\Gamma$ point.  It is noted that these bands are well separated from each other around the {\it X} point due to the absence of band inversion [\onlinecite{Oinuma2017}]. On the other hand, in LaBi [Fig. 1(c)], the La 5{\it d} and Bi 6{\it p} bands cross each other midway between the {\it X} and $\Gamma$ points (marked by white arrows), in support of the inverted band structure. We found that such a band inversion is also resolved in CeBi consistent with the previous study [\onlinecite{Kuroda2018}], although there exist some quantitative differences between CeBi and LaBi, {\it e.g.}, the Bi 6{\it p} band around $E_{\rm F}$ is flatter in CeBi than in LaBi, leading to the movement of the band-crossing point toward the $\Gamma$ point in CeBi.
 
 \begin{figure}
\includegraphics[width=2.8 in]{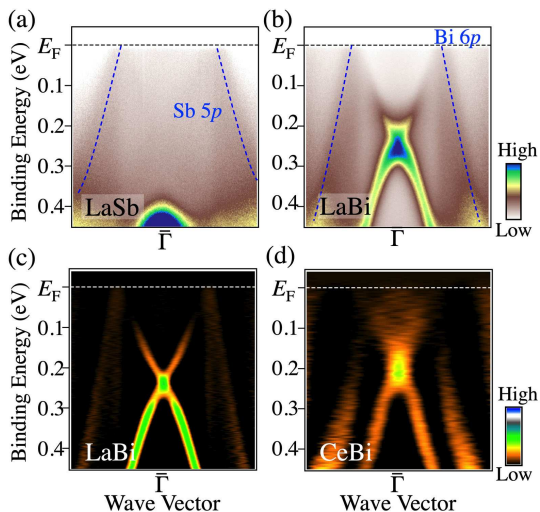}
\hspace{0.2in}
\caption{(color online).(a), (b) ARPES intensity of LaSb and LaBi, respectively, measured around the $\bar{\Gamma}$ point at {\it T} = 30 K with the Xe-I line ($h\nu$ = 8.437 eV). (c), (d) Second-derivative-intensity plot of MDCs for LaBi and CeBi, respectively.
}
\end{figure}

 To clarify a possible link between the bulk-band inversion and the appearance of topological surface states (SS), we have performed high-resolution ARPES measurements with VUV photons (the Xe-I line; $h\nu$ = 8.437 eV) around the $\bar{\Gamma}$ point where a distinct Dirac-cone-like energy band was resolved in previous ARPES studies on LaBi and CeBi [\onlinecite{WuPRB2016, NiuPRB2016, NayakNC2017, LouPRB2017, Kuroda2018, BFeng2018, PLi2018}]. As seen from the ARPES-intensity plot of LaSb [Fig. 2(a)], no bands are observed inside the inner Sb 5{\it p} band crossing $E_{\rm F}$. In a sharp contrast, in LaBi, there exists an X-shaped Dirac-cone-like band besides the holelike Bi 6{\it p} band [Fig. 2(b)]. This is better visualized in the second-derivative-intensity plot of momentum distribution curves (MDCs) shown in Fig. 2(c). Such a crucial difference between LaSb and LaBi strongly suggests that the band inversion and the resultant change in the bulk-band topology are indeed related to the appearance of the Dirac-cone SS. This is also collaborated with the observation of a similar Dirac-cone SS in the paramagnetic phase ({\it T} = 30 K) in CeBi, as displayed in Fig. 2(d) (note that the spectral feature of CeBi is broader due to the poorer surface quality).
 
 Now we proceed to our most important finding, a striking temperature dependence of the topological SS in CeBi. We systematically performed a temperature-dependent ARPES measurement across the two-step AF transition at $T_{\rm AFI}$ (= 25 K) and $T_{\rm AFII}$ (= 14 K) [see Fig. 3(c) for the corresponding magnetic structure [\onlinecite{RosMig3}, \onlinecite{Kohgi2000}]]. As shown in Fig. 3(a), the energy distribution curve (EDC) at the $\bar{\Gamma}$ point in the paramagnetic phase ({\it T} = 30 K) consists of a single peak around the Dirac-point energy at $\sim$0.2 eV [see Fig. 2(d)]. On lowering temperature, two peaks suddenly appear at $E_{\rm B}$ $\sim$ 0.14 and $\sim$ 0.27 eV, respectively, at around {\it T} = 24 K ($\sim$$T_{\rm AFI}$), but they are transformed again into a single peak below {\it T} = 13 K ($\sim$$T_{\rm AFI}$) at $\sim$0.2 eV with an additional faint feature at $\sim$0.1 eV. Such behavior is better visualized in the second-derivative-intensity plot of EDC in Fig. 3(b), where one can recognize that the transition of spectral feature is abrupt and apparently linked to the AF transition.
 
 To gain further insight into the spectral change related to the AF transition, we show in Figs. 4(a)-4(c) the ARPES-derived band dispersions at representative temperatures in the paramagnetic (30 K), AF-I (19 K), and AF-II (6 K) phases, respectively, obtained by taking the second derivative of MDCs. While one can recognize a single Dirac-cone band at {\it T} = 30 K in the paramagnetic phase [Fig. 4(a)], there exit two Dirac-cone-like bands at {\it T} = 19 K in the AF-I phase [Fig. 4(b)] which are energetically separated from each other by $\sim$0.13 eV; this value corresponds to the energy separation between the two peaks in the EDC in Fig. 3(a) (note that the band dispersion of the lower Dirac-cone branch above the Dirac point is not so clearly visible). Surprisingly, at {\it T} = 6 K in the AF-II phase [Fig. 4(c)], the two Dirac-cone bands again disappear and a single Dirac-cone band recovers. In contrast, the holelike Bi-6{\it p} bulk band shows no discernible band reconstruction even across $T_{\rm AFII}$ (note that this band is gradually pushed up on lowering temperature likely due to {\it p-f} mixing as reported in CeSb [\onlinecite{Kumi1997, Ito2004, Takayama2009, SJang2019}]; see Appendix A for detailed temperature dependence of the bulk bands). To the best of our knowledge, the present result is the first experimental observation of the antiferromagnetism-induced drastic reconstruction of the Dirac-cone SS.
 
 \begin{figure}
 \includegraphics[width=3 in]{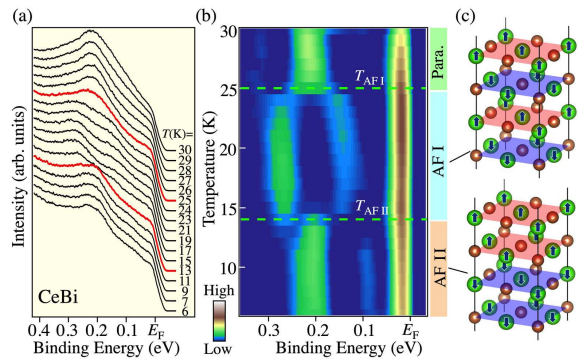}
 \hspace{0.2in}
\caption{(color online). (a) Temperature dependence of EDC at the $\bar{\Gamma}$ point for CeBi measured with the Xe-I photons ($h\nu$= 8.437 eV) across the two-step magnetic transition at 25 and 14 K. EDCs at around $T_{\rm AFI}$ and $T_{\rm AFII}$ are highlighted by red color. (b) Second-derivative-intensity plot of EDC at the $\bar{\Gamma}$ point as a function of temperature. Note that the temperature-independent flat feature near $E_{\rm F}$ is due to the Fermi-edge cut-off. (c) Schematic view of magnetic structure in the (top) AF-I and (bottom) AF-II phases (the case for type-F domain).
}
\end{figure}
 
 \begin{figure}
\includegraphics[width=2.6in]{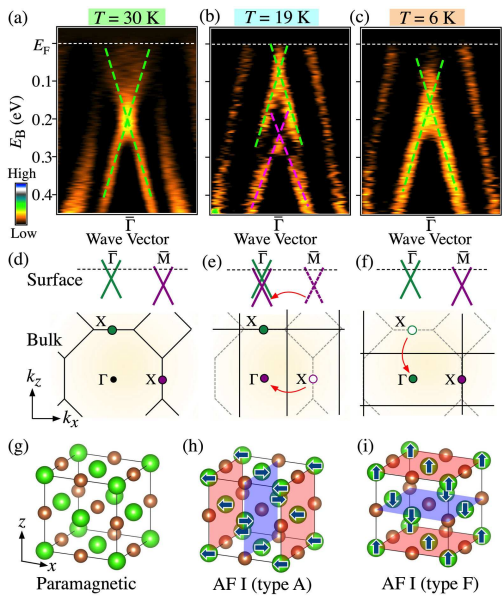}
\hspace{0.2in}
\caption{(color online). (a)-(c) Plot of second-derivative intensity of MDCs around the $\bar{\Gamma}$ point for CeBi measured at {\it T} = 30 K (paramagnetic phase), 19 K (AF-I phase), and 6 K (AF-II phase). (d)-(f) Schematic view of band folding for the Dirac-cone SS in the $k_x$-$k_z$ plane. (g)-(i) Crystal and magnetic structures for paramagnetic, AF-I type-A domain, and AF-I type-F domain, respectively [\onlinecite{RosMig3}, \onlinecite{Kohgi2000}]. Red and blue shades highlight the ferromagnetic Ce layer.
}
\end{figure}

 Now we discuss the origin of observed intriguing change in the Dirac-cone band dispersion. The appearance of a single Dirac-cone band at the  $\bar{\Gamma}$ point in the paramagnetic phase is reasonable since a single bulk {\it X} point is projected onto the surface $\bar{\Gamma}$ point as shown by a schematic band diagram in the $k_x$-$k_z$ plane in Fig. 4(d). In this case, the band inversion takes place just once, resulting in a single Dirac-cone band at the $\bar{\Gamma}$ point. In the AF-I phase, the magnetic moment of Ce 4{\it f} electron aligns ferromagnetically in a single Ce layer as shown by red or blue shade in Fig. 4(h), but aligns antiferromagnetically between adjacent layers (compare blue and red shades). Under zero magnetic field, there exist three types of magnetic domains, two type-A domains with a ferromagnetic layer perpendicular to the surface [one of two such cases is shown in Fig. 4(h)], and a type-F domain with a ferromagnetic layer oriented parallel to the surface [Fig. 4(i)]. Such AF domains give rise to a superstructure potential of 2$\times$1$\times$1 (1$\times$2$\times$1 and 1$\times$1$\times$2 as well). In such a case, it is expected that the original band is folded with respect to the magnetic BZ. As shown in Fig. 4(e), since the 2$\times$1$\times$1 magnetic BZ boundary (black solid line) is located at the midpoint between the  $\bar{\Gamma}$ and $\bar{M}$ points in the surface BZ, it is expected that the Dirac cone originally located at the $\bar{M}$ point is folded onto the $\bar{\Gamma}$ point, and {\it vice versa}. Thus, one of double Dirac cones seen in Fig. 4(b) may be attributed to the Dirac cone originally located at the $\bar{M}$ point in the paramagnetic phase. Such a Dirac cone at $\bar{M}$ was identified in previous VUV-ARPES studies on LaBi and CeBi [\onlinecite{NiuPRB2016, NayakNC2017, LouPRB2017, Kuroda2018, BFeng2018, PLi2018}], although the shape is more complex than that at the  $\bar{\Gamma}$ point probably because two bulk {\it X} points are simultaneously projected onto the surface $\bar{M}$ point [\onlinecite{NiuPRB2016}] (note that the $\bar{M}$ point is not accessible in our ARPES measurement due to the insufficient $k$ range of Xe-I photons). In this context, the observed double Dirac-cone feature is not fully explained in terms of a simple overlap of the Dirac cones at $\bar{\Gamma}$ and $\bar{M}$ in the paramagnetic phase, and one may need to invoke a complex mechanism beyond the simple band-folding picture (see Appendix B for more detailed discussion of the band folding). Thus, the band-folding picture is just a likely possibility and its verification requires high-resolution domain-selective measurements.

Here we point out some other possibilities to explain the emergence of a double Dirac-cone feature in the AF-I phase. We think that this Dirac-cone-like feature is hardly explained in terms of the folded bulk band. Because, the $k_z$ point in the present measurement  ($k_z$ = $0.24\pi$) estimated from the inner-potential value ($V_0$ = 13.5 eV) is far from the bulk {\it X} point ($k_z$ = 0 or $\pi$). Thus, in the present experimental setup, the electronlike bulk band should be located far above $E_{\rm F}$ after folding and  as a result unable to produce the observed Dirac-cone feature near $E_{\rm F}$  even after band-folding. Moreover, the bulk band in the paramagnetic phase does not show a Dirac-cone-like dispersion at the {\it X} point. This is obviously incompatible with the bulk origin of the double Dirac cone. One may also think that one of the Dirac-cone bands in the AF-I phase could be a trivial SS which is seen in the paramagnetic phase of LaSb [\onlinecite{NiuPRB2016}]. However, such trivial SS is not well visible in the present study of LaSb [Fig. 2(a)], probably because of the difference in the experimental conditions (such as photon energy, light polarization, and sample geometry). Since we used the same experimental conditions for LaSb and CeBi in our ARPES study, the trivial SS would have shown up even in the paramagnetic phase in CeBi if some of observed bands in the AF phase are attributed to the trivial SS. However, this is not the case in our experiment since we resolve a single non-trivial SS (a Dirac cone) in the paramagnetic phase of CeBi, as shown in Figs. 2(d) and 4(a). It is noted that the observed double Dirac-cone feature may originate from an overlap of SS in the type-F and type-A domains which are energetically inequivalent to each other. One cannot rule out this possibility because the $\bar{M}$ point is out of the measurable $k$ range with a Xe lamp. To resolve this issue, it is highly desirable to perform a high-resolution domain-selective ARPES measurements covering both the $\bar{\Gamma}$ and $\bar{M}$ points to experimentally distinguish the SS originating from the type-A and type-F domains.

One may point out that the gapless topological SS should not appear in the AF phase since the topological properties can be destroyed by the antiferromagnetism {\it via} time-reversal-symmetry (TRS) breaking. However, we think that the topological SS can be maintained even in the AF phase when we take into account a combined symmetry. It has been predicted that even antiferromagnets, where both TRS ($\Theta$) and primitive lattice translational symmetry ($T_{\rm 1/2}$) are broken, can still host a topological phase if the crystal preserves the combined symmetry $S = \Theta T_{\rm 1/2}$ [\onlinecite{MongPRB2010}]. This is likely the case for the type-A domain in the AF-I phase because the AF vector satisfies this translation condition. This may be the reason why the Dirac-cone SS is still observed in the AF-I phase. It is noted that, since the combined symmetry is broken at the surface of the type-F domain, the Dirac-cone SS is expected to exhibit a finite energy gap at the Dirac point. However, this gap would be very small and difficult to resolve with the present experimental resolution.

 The appearance of a single Dirac cone in the AF-II phase [Fig. 4(c)] as opposed to the double Dirac cone in the AF-I phase is puzzling. The AF domain in the AF-II phase creates the 4$\times$1$\times$1 potential [see Fig. 3(c)] and gives rise to three magnetic BZ boundaries between $\bar{\Gamma}$ and $\bar{M}$, resulting in the emergence of several Dirac cones between $\bar{\Gamma}$ and $\bar{M}$. However, this is not experimentally observed since such multiple Dirac cones at corresponding $k$ regions are not resolved in the present ARPES experiment. We speculate that the surface electrons do not strongly feel the periodic potential with such a long periodicity so that the influence of the band folding would be weakened compared in the AF-I phase with a shorter periodic potential. But we point out here that the above interpretation based on the in-plane band-folding is one of possibilities. Moreover, there exist some unresolved issues as to (i) why the bulk bands do not show clear band folding in the AF phase unlike the SS, and (ii) whether the electronic states at $\bar{M}$ in the AF phase can be explained by the band-folding picture. The verification requires the domain-selective high-resolution measurement by micro-beam-spot ARPES, combined with the sophisticated first-principles band-structure calculations that take into account the magnetic structure.
 
\section{SUMMARY}
The present ARPES study of CeBi has revealed the existence of topological Dirac-cone SS at the $\bar{\Gamma}$ point associated with the band inversion at the {\it X} point of bulk BZ. We uncovered an unexpected change in the energy dispersion of the Dirac-cone SS associated with the two-step AF transition. Intriguingly, we found that a single Dirac-cone band observed in the paramagnetic phase abruptly becomes a double Dirac cone on entering the AF-I phase below 25 K, whereas the doubling disappears in the AF-II phase below 14 K. The present result strongly suggests a crucial role of antiferromagnetism to the appearance of the Dirac-cone SS, and opens a pathway toward understanding the interplay between magnetism and topology in exotic topological materials.

\begin{acknowledgments}
We thank K. Sugawara, M. Kitamura, R. Yukawa, S. Ideta, and K. Tanaka and for their assistance in the ARPES experiments. This work was supported by JST-CREST (No: JPMJCR18T1), JST-PRESTO (No: JPMJPR18L7), MEXT of Japan (Innovative Area ``Topological Materials Science" JP15H05853), JSPS (JSPS KAKENHI No: JP17H01139, JP26287071, JP25220708, JP18J20058, JP18H01160), KEK-PF (Proposal number: 2018S2-001, 2018G653), and UVSOR (Proposal number: 30-858, 30-846, 30-554, 30-568).\end{acknowledgments}

\appendix
\section{TEMPERATURE DEPENDENCE OF BULK-BAND DISPERSION}

 We have examined the influence of {\it p-f} mixing on the observed electronic structure, by inspecting the change in the bulk valence-band dispersion with temperature. As shown in Figs. 5(a) and 5(b), one can recognize in the paramagnetic phase ({\it T} = 30 K) a couple of holelike Bi-6{\it p} bands (inner and outer) which cross $E_{\rm F}$ besides the surface band. Interestingly, the magnitude of the Fermi wave vectors ($k_{\rm F}$) of these bands increases in the AF-II phase, as shown by white arrows. To investigate the detailed temperature evolution of the $k_{\rm F}$ position, we show in Fig. 5(c) the second-derivative intensity of MDCs at $E_{\rm F}$ as a function of temperature. One can clearly see a systematic increase in the absolute value of $k_{\rm F}$ on lowering temperature. A similar behavior in the valence band has been observed in CeSb [\onlinecite{Kumi1997, Ito2004, Takayama2009, SJang2019}] and is interpreted in terms of the energy shift of valence bands due to the {\it p-f} mixing. Thus, our experimental results clearly show that the {\it p-f} mixing affects the band structure also in CeBi.

\begin{figure}[h]
\includegraphics[width=3.2 in]{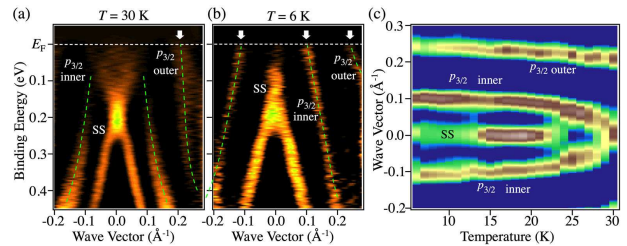}
\hspace{0.2in}
\caption{(a), (b) Plots of second-derivative intensity of MDCs around the $\bar{\Gamma}$ point for CeBi measured at {\it T} = 30 K (paramagnetic phase) and 6 K (AF-II phase) [same as Figs. 4(a) and 4(c)]. Green dashed curves are a guide for the eyes to trace the holelike Bi 6$p_{3/2}$ bulk bands. Arrows indicate the position of the $k_{\rm F}$ points for the bulk bands. (c) Temperature dependence of the second-derivative intensity of MDCs at $E_{\rm F}$.
}
\end{figure}

\section{ANALYSIS OF EDCs and MDCs}

\begin{figure}
\includegraphics[width=2.6in]{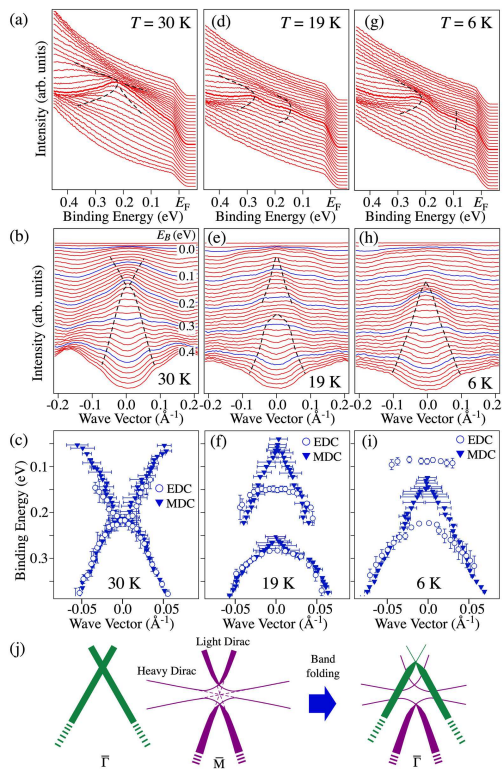}
\hspace{0.2in}
\caption{(a), (b) Near-$E_{\rm F}$ EDCs and MDCs of CeBi, respectively, measured along the $\overline{\Gamma}\overline{M}$ cut at {\it T} = 30 K (paramagnetic phase). (c) Comparison of band energies estimated from the peak position of the EDCs and MDCs in (a) and (b). (d)-(f) Same as (a)-(c) but for the AF-I phase ({\it T} = 19 K). (g)-(i) Same as (a)-(c) but for the AF-II phase ({\it T} = 6 K). (j) Schematic band-folding picture in the AF-I phase which takes into account the folding of light and heavy Dirac cones around the $\bar{M}$ point. Note that the heavy Dirac-cone band is not clearly seen in the present study.
}
\end{figure}

 We show in Figs. 6(a)-6(i) the EDCs, MDCs, and their numerical analyses, at representative temperatures in the paramagnetic phase ({\it T} = 30 K), the AF-I phase ({\it T} = 19 K), and the AF-II phase ({\it T} = 6 K). As shown in Figs. 6(a) and 6(b), one can see in the paramagnetic phase ({\it T} = 30 K) a single X-shaped band in both EDCs and MDCs. We traced the peak position of the EDCs and MDCs by numerical fittings with Lorentzians around the peak top, and found that the obtained band energies for EDCs and MDCs reasonably coincide with each other at {\it T} = 30 K near the Dirac point, as shown in Fig. 6(c). In the AF-I phase [Fig. 6(f)], on the other hand, the EDC- and MDC-peaks show a finite deviation around the $\bar{\Gamma}$ point, whereas they reasonably overlap with each other away from the $\bar{\Gamma}$ point. This suggests that the lower Dirac-cone bands are substantially rounded around the Dirac point. The rounded behavior of the Dirac-cone dispersion may be explained in terms of a finite gap-opening at the Dirac point and/or the inherent curvature of the lower Dirac-cone band in the AF phase due to the complicated hybridization between the original and folded bands.
 
 Here we elaborate on our observation of the double Dirac-cone feature based on the band-folding picture. It has been reported that the SS at the $\bar{M}$ point in the surface Brillouin zone (corresponding to the {\it X} point in bulk) is characterized by light and heavy Dirac-cone bands [\onlinecite{Kuroda2018}, \onlinecite{PLi2018}] with an energy gap at the Dirac point due to hybridization, as schematically shown in Fig. 6(j). In our experiment, we set the analyzer slit parallel to the $\overline{\Gamma}\overline{M}$ direction. In this geometry, when the bands at the $\bar{M}$ point are folded onto the $\bar{\Gamma}$ point due to antiferromagnetism, we expect to observe a narrow energy dispersion of the heavy Dirac-cone band besides the light one around the $\bar{\Gamma}$ point in Fig. 4(b). However, we have not resolved such a heavy Dirac-cone band,  presumably because of the matrix-element effect of photoelectron intensity. As illustrated in Fig. 6(j), the band which is topped at $E_{\rm B} \sim$ 0.25 eV in the EDCs likely corresponds to the lower branch of the gapped light Dirac cone since the energy position and the rounded shape are similar to those observed in the paramagnetic phase in CeBi (see ref. [\onlinecite{Kuroda2018}]). On the other hand, the band topped at $\sim$0.06 eV in the EDC would likely be the original Dirac cone at the $\bar{\Gamma}$ point which was pushed upward relative to the paramagnetic one due to the hybridization with the folded band. The upper branch of the light Dirac band was not clearly resolved in the present study probably because it largely overlaps with the lower branch of the original Dirac cone. However, this interpretation is just one of possible scenarios at present and needs to be verified by the domain-selective measurements.

As shown in the EDCs for the AF-II phase in Fig. 6(g), there exit two peaks at  $E_{\rm B} \sim$ 0.2 and 0.1 eV at the $\bar{\Gamma}$ point. The 0.2-eV peak has a holelike dispersion and disperses toward higher $E_{\rm B}$ on moving away from the $\bar{\Gamma}$ point. This feature is also seen in the MDCs in Figs. R2(h) and R2(i) as a holelike band which has a top at $E_{\rm B} \sim$ 0.15 eV. We also found that the 0.1-eV peak in the EDCs is nearly flat around the $\bar{\Gamma}$ point and appears to slightly disperse toward higher $E_{\rm B}$ on moving away from the $\bar{\Gamma}$ point while rapidly reducing its intensity. This holelike dispersion is hardly explained in terms of a gapped upper Dirac-cone band because the upper Dirac cone should have an electronlike dispersion. The reason why the energy position of the Dirac-crossing point in the MDCs and the peaks in the EDCs does not well coincide with each other [see Fig. 6(i)] may be due to the deviation of the Dirac-cone-like band from a linear dispersion, as well as the unusually weak intensity of the 0.1-eV peak away from the $\bar{\Gamma}$ point (which makes it difficult to follow its dispersion from the MDC plots). One may argue that the existence of a couple of holelike bands in the AF-II phase can be explained in terms of a double Dirac-cone band similar to that in the AF-I phase but shifted toward $E_{\rm F}$ due to the {\it p-f} mixing. However, whereas the bulk holelike band certainly shows a gradual shift of the dispersion toward  $E_{\rm F}$ on lowering temperature due to the {\it p-f} mixing [see Fig. 5(c)], the observed abrupt shift of the band energy across $T_{\rm AFII}$ cannot be well explained in terms of the {\it p-f} mixing, since such an abrupt shift is absent in the bulk bands.

\bibliographystyle{prsty}

\begin{thebibliography}{60}
\bibitem{RosMig1} J. Rossat-Mignod, P. Burlet, J. Villain, H. Bartholin, W. Tcheng-Si, D. Florence, and O. Vogt, Phys. Rev. B \textbf{16}, 440 (1977).

\bibitem{RosMig2} J. Rossat-Mignod, P. Burlet, S. Quezel, J. M. Effantin, D. Delac\^{o}te, H. Bartholin, O. Vogt, and D. Ravot, J. Magn. Magn. Mater.  \textbf{31-34}, 398 (1983).

\bibitem{RosMig3} H. Bartholin, P. Burlet, S. Quezel, J. Rossat-Mignod, and O. Vogt, Le J. Phys. Colloq. \textbf{40}, C5 (1979).

\bibitem{Kohgi2000} M. Kohgi, K. Iwasa, and T. Osakabe, Physica B \textbf{281-282}, 417 (2000).

\bibitem{Hasegawa} A. Hasegawa, J. Phys. Soc. Jpn.  \textbf{54}, 677 (1985).

\bibitem{Settai1994} R. Settai, T. Goto, S. Sakatume, Y. S. Kwon, T. Suzuki, Y. Kaneta, and O. Sakai, J. Phys. Soc. Jpn.  \textbf{63}, 3026 (1994).

\bibitem{Kasuya_pf} H. Takahashi and T. Kasuya, J. Phys. C  \textbf{18}, 2697 (1985);  \textbf{18}, 2709 (1985);  \textbf{18}, 2721 (1985);  \textbf{18}, 2731 (1985);  \textbf{18}, 2745 (1985);  \textbf{18}, 2755 (1985).

\bibitem{CF} H. Heer, A. Furrer, W. Halg and O. Vogt, J. Phys. C \textbf{12}, 5207 (1979).

\bibitem{Kumi1997} H. Kumigashira, H.-D. Kim, A. Ashihara, A. Chainani, T. Yokoya, T. Takahashi, A. Uesawa, and T. Suzuki, Phys. Rev. B \textbf{56}, 13654 (1997).

\bibitem{Ito2004} T. Ito, S. Kimura, and H. Kitazawa, Physica B  \textbf{351}, 268 (2004).

\bibitem{Takayama2009} A. Takayama, S. Souma, T. Sato, T. Arakane, and T. Takahashi, J. Phys. Soc. Jpn. \textbf{78}, 073702 (2009).

\bibitem{SJang2019} S. Jang, R. Kealhofer, C. John, S. Doyle, J. Hong, J. H. Shim, Q. Si, O. Erten, J. D. Denlinger, and J. G. Analytis, Sci. Adv. \textbf{5}, eaat7158 (2019).

\bibitem{Kitazawa1988}H. Kitazawa, Y. S. Kwon, A. Oyamada, N. Takeda, H. Suzuki, S. Sakatsume, T. Satoh, T. Suzuki, and T. Kasuya, J. Magn. Magn. Mater. \textbf{76-77}, 40 (1988).

\bibitem{Boehm1979} J. von Boehm and P. Bak, Phys. Rev. Lett. \textbf{42}, 122 (1979).

\bibitem{Nakanishi1989} K. Nakanishi, J. Phys. Soc. Jpn. \textbf{58}, 1296 (1989).

\bibitem{Kasuya1990} T. Kasuya, Y. S. Kwon, T. Suzuki, K. Nakanishi, F. Ishiyama, and K. Takegahara, J. Magn. Magn. Mater. \textbf{90-91}, 389 (1990).

\bibitem{Okayama1992} Y. Okayama, H. Takahashi, N. Mori, Y. S. Kwon, Y. Haga, and T. Suzuki, J. Magn. Magn. Mater. \textbf{108}, 113 (1992).

\bibitem{Kasuya1993} T. Kasuya, Y. Haga, Y. S. Kwon, and T. Suzuki, Physica B \textbf{186-188}, 9 (1993).

\bibitem{Iwasa2002} K. Iwasa, A. Hannan, M. Kohgi, and T. Suzuki, Phys. Rev. Lett. \textbf{88}, 207201 (2002).

\bibitem{ZengArXiv} M. Zeng, C. Fang, G. Chang, Y.-A. Chen, T. Hsieh, A. Bansil, H. Lin, and L. Fu, arXiv:1504.03492 (2015).

\bibitem{PJGuo2016} P.-J. Guo, H.-C. Yang, B.-J. Zhang, K. Liu, and Z.-Y. Lu, Phys. Rev. B \textbf{93}, 235142 (2016).

\bibitem{PJGuo2017} P.-J. Guo, H.-C. Yang, K. Liu, and Z. Y. Lu, Phys. Rev. B \textbf{96}, 081112(R) (2017).

\bibitem{Stepanov2015} N. N. Stepanov, N. V. Morozova, A. E. Kar'kin, A. V. Golubkov, and V. V. Kaminskii, Phys. Solid State \textbf{57}, 2369 (2015). 

\bibitem{TaftiNP2016} F. F. Tafti, Q. D. Gibson, S. K. Kushwaha, N. Haldolaarachchige, and R. J. Cava, Nat. Phys. \textbf{12}, 272 (2016).

\bibitem{SunNJP2016} S. S. Sun, Q. Wang, P. J. Guo, K. Liu, and H. C. Lei, New J. Phys. \textbf{18}, 082002 (2016).

\bibitem{KumarPRB2016} N. Kumar, C. Shekhar, S.-C. Wu, I. Leermakers, O. Young, U. Zeitler, B. H. Yan, and C. Felser, Phys. Rev. B \textbf{93}, 241106(R) (2016).

\bibitem{TaftiPNAS2016} F. F. Tafti, Q. D. Gibson, S. K. Kushwaha, J. W. Krizan, N. Haldolaarachchige, and R. J. Cava, Proc. Natl. Acad. Sci. \textbf{113}, E3475-E3481 (2016).

\bibitem{Kumar2017} N. Kumar, C. Shekhar, J. Klotz, J. Wosnitza, and C. Felser, Phys. Rev. B \textbf{96}, 161103(R) (2017).

\bibitem{TaftiPRB2017} F. F. Tafti, M. S. Torikachvili, R. L. Stillwell, B. Baer, E. Stavrou, S. T. Weir, Y. K. Vohra, H.-Y. Yang, E. F. McDonnell, S. K. Kushwaha, Q. D. Gibson, R. J. Cava, and J. R. Jeffries, Phys. Rev. B \textbf{95}, 014507 (2017).

\bibitem{SinghaarXiv2017} R. Singha, B. Satpati, and P. Mandal, arXiv:1703.06100v1.

\bibitem{FWu2017} F. Wu, C. Y. Guo, M. Smidman, J. L. Zhang, and H. Q. Yuan, Phys. Rev. B \textbf{96}, 125122 (2017).

\bibitem{CGuo2017} C. Guo, C. Cao, M. Smidman, F. Wu, Y. Zhang, F. Steglich, F. C. Zhang, and H. Yuan, Npj Quantum Mater. \textbf{2}, 39 (2017).

\bibitem{Joe2018} L. Ye, T. Suzuki, C. R. Wicker, and J. G. Checkelsky, Phys. Rev. B \textbf{97}, 081108(R) (2018).

\bibitem{YWang2018B} Y.-Y. Wang, L.-L. Sun, S. Xu, Y. Su, and T.-L. Xia, Phys. Rev. B \textbf{98}, 045137 (2018).

\bibitem{YWang2018A} Y.-Y. Wang, H. Zhang, X.-Q. Lu, L.-L. Sun, S. Xu, Z.-Y. Lu, K. Liu, S. Zhou, and T.-L. Xia, Phys. Rev. B \textbf{97}, 085137 (2018).

\bibitem{ZLi2017} Z. Li, D.-D. Xu, S.-Y. Ning, H. Su, T. Iitaka, T. Tohyama, and J.-X. Zhang, Int. J. Mod. Phys. B \textbf{31}, 1750217 (2017).

\bibitem{ZengPRL2016} L.-K. Zeng, R. Lou, D.-S. Wu, Q. N. Xu, P.-J. Guo, L.-Y. Kong, Y.-G. Zhong, J.-Z. Ma, B.-B. Fu, P. Richard, P. Wang, G. T. Liu, L. Lu, Y.-B. Huang, C. Fang, S.-S. Sun, Q. Wang, L. Wang, Y.-G. Shi, H. M. Weng, H.-C. Lei, K. Liu, S.-C. Wang, T. Qian, J.-L. Luo, and H. Ding, Phys. Rev. Lett. \textbf{117}, 127204 (2016).

\bibitem{WuPRB2016} Y. Wu, T. Kong, L.-L. Wang, D. D. Johnson, D. Mou, L. Huang, B. Schrunk, S. L. Bud'ko, P. C. Canfield, and A. Kaminski, Phys. Rev. B \textbf{94}, 081108(R) (2016).

\bibitem{HasanArxiv2016} N. Alidoust, A. Alexandradinata, S.-Y. Xu, I. Belopolski, S. K. Kushwaha, M. Zeng, M. Neupane, G. Bian, C. Liu, D. S. Sanchez, P. P. Shibayev, H. Zheng, L. Fu, A. Bansil, H. Lin, R. J. Cava, and M. Z. Hasan, arXiv:1604.08571v1.

\bibitem{NiuPRB2016} X. H. Niu, D. F. Xu, Y. H. Bai, Q. Song, X. P. Shen, B. P. Xie, Z. Sun, Y. B. Huang, D. C. Peets, and D. L. Feng, Phys. Rev. B \textbf{94}, 165163 (2016).

\bibitem{JHe2016} J. He, C. Zhang, N. J. Ghimire, T. Liang, C. Jia, J. Jiang, S. Tang, S. Chen, Y. He, S.-K. Mo, C. C. Hwang, M. Hashimoto, D. H. Lu, B. Moritz, T. P. Devereaux, Y. L. Chen, J. F. Mitchell, and Z.-X. Shen, Phys. Rev. Lett. \textbf{117}, 267201 (2016).

\bibitem{NayakNC2017} J. Nayak, S.-C. Wu, N. Kumar, C. Shekhar, S. Singh, J. Fink, E. E. D. Rienks, G. H. Fecher, S. S. P. Parkin, B. Yan, and C. Felser, Nat. Commun. \textbf{8}, 13942 (2017).

\bibitem{Oinuma2017}H. Oinuma, S. Souma, D. Takane, T. Nakamura, K. Nakayama, T. Mitsuhashi, K. Horiba, H. Kumigashira, M. Yoshida, A. Ochiai, T. Takahashi, and T. Sato, Phys. Rev. B \textbf{96}, 041120(R) (2017).

\bibitem{LouPRB2017} R. Lou, B.-B. Fu, Q. N. Xu, P.-J. Guo, L.-Y. Kong, L.-K. Zeng, J.-Z. Ma, P. Richard, C. Fang, Y.-B. Huang, S.-S. Sun, Q. Wang, L. Wang, Y.-G. Shi, H. C. Lei, K. Liu, H. M. Weng, T. Qian, H. Ding, and S.-C. Wang, Phys. Rev. B \textbf{95}, 115140 (2017).

\bibitem{YWu2017}Y. Wu, Y. Lee, T. Kong, D. Mou, R. Jiang, L. Huang, S. L. Bud'ko, P. C. Canfield, and A. Kaminski, Phys. Rev. B \textbf{96}, 035134 (2017).

\bibitem{Kuroda2018}	K. Kuroda, M. Ochi, H. S. Suzuki, M. Hirayama, M. Nakayama, R. Noguchi, C. Bareille, S. Akebi, S. Kunisada, T. Muro, M. D. Watson, H. Kitazawa, Y. Haga, T. K. Kim, M. Hoesch, S. Shin, R. Arita, and T. Kondo, Phys. Rev. Lett. \textbf{120}, 086402 (2018).

\bibitem{BFeng2018} B. Feng, J. Cao, M. Yang, Y. Feng, S. Wu, B. Fu, M. Arita, K. Miyamoto, S. He, K. Shimada, Y. Shi, T. Okuda, and Y. Yao, Phys. Rev. B \textbf{97}, 155153 (2018).

\bibitem{PLi2018}P. Li, Z. Wu, F. Wu, C. Cao, C. Guo, Y. Wu, Y. Liu, Z. Sun, C.-M. Cheng, D.-S. Lin, F. Steglich, H. Yuan, T.-C. Chiang, and Y. Liu, Phys. Rev. B \textbf{98}, 085103 (2018).

\bibitem{MongPRB2010} R. S. K. Mong, A. M. Essin, and J. E. Moore, Phys. Rev. B \textbf{81}, 245209 (2010).

\bibitem{HGuo2011} H. Guo, S. Feng, and S.-Q. Shen, Phys. Rev. B \textbf{83}, 045114 (2011).

\bibitem{CFang2013} C. Fang, M. J. Gilbert, and B. A. Bernevig, Phys. Rev. B \textbf{88}, 085406(R) (2013).

\bibitem{Miyakoshi2013} S. Miyakoshi and Y. Ohta, Phys. Rev. B \textbf{87}, 195133 (2013).

\bibitem{Yoshida2013} T. Yoshida, R. Peters, S. Fujimoto, and N. Kawakami, Phys. Rev. B \textbf{87}, 085134 (2013).

\bibitem{Muller2014} R. A. M\"{u}ller, N. R. Lee-Hone, L. Lapointe, D. H. Ryan, T. Pereg-Barnea, A. D. Bianchi, Y. Mozharivskyj, and R. Flacau, Phys. Rev. B. \textbf{90}, 041109(R) (2014).

\bibitem{CXLiu2014} C.-X. Liu, R.-X. Zhang, and B. K. VanLeeuwen, Phys. Rev. B \textbf{90}, 085304 (2014).

\bibitem{RXZhang2015} R.-X. Zhang and C.-X. Liu, Phys. Rev. B \textbf{91}, 115317 (2015).

\bibitem{CFang2015} C. Fang and L. Fu, Phys. Rev. B \textbf{91}, 161105(R) (2015).

\bibitem{JYu2017} J. Yu, B. Yan, and C.-X. Liu, Phys. Rev. B \textbf{95}, 235158 (2017).

\bibitem{Brzezicki2017} W. Brzezicki and M. Cuoco, Phys. Rev. B \textbf{95}, 155108 (2017).

\bibitem{NHao2017} N. Hao, F. Zheng, P. Zhang, and S.-Q. Shen, Phys. Rev. B \textbf{96}, 165102 (2017).

\bibitem{KWChang2018} K.-W. Chang and P.-J. Chen, Phys. Rev. B \textbf{9}7, 195145 (2018).

\bibitem{Franciosi1981} A. Franciosi, J. H. Weaver, N. M{\aa}rtensson, and M. Croft, Phys. Rev. B \textbf{24}, 3651 (1981).

\bibitem{Allen1981} J. W. Allen, S. J. Oh, I. Lindau, J. M. Lawrence, L. I. Johansson, and S. B. Hagstr\"{o}m, Phys. Rev. Lett. \textbf{46}, 1100 (1981).


 
\end{thebibliography}

\end{document}